\documentclass[a4paper,12pt,BCOR5mm,bibliography=totoc]{article}

\usepackage{caption}
\usepackage{subcaption}

\usepackage[english]{babel}
\usepackage{fancyhdr}
\usepackage{setspace}
\usepackage{graphics}
\usepackage{graphicx}
\usepackage{eso-pic}
\usepackage{amsmath}
\usepackage{amstext}
\usepackage{amsfonts}
\usepackage{amssymb}
\usepackage{hyperref}
\usepackage{slashed}
\usepackage{cite}
\usepackage{bbold}
\usepackage{multirow}
\usepackage{esdiff}
\usepackage[font=small,labelfont=small]{caption}
\usepackage{commath}
\usepackage{tikz-cd}
\usepackage{mathabx}
\usepackage{relsize}
\usepackage{simplewick} 

\usepackage[affil-it]{authblk} 
 
\renewcommand{\vec}[1]{\ensuremath{\boldsymbol{#1}}}

\newcommand{\iu}{\mathrm{i}}	
\newcommand{\ee}{\mathrm{e}}	

\newcommand{\bra}[1]{\ensuremath{\left< #1\,\right|}}
\newcommand{\ket}[1]{\ensuremath{\left|\, #1\right>}}

\begin{document}

\vspace{0.01cm}
\begin{center}
{\Large\bf  Black Hole Based Quantum Computing in Labs and in the Sky} 

\end{center}

\vspace{0.1cm}

\begin{center}

{\bf Gia Dvali}$^{a,b,c}$ and {\bf Mischa Panchenko}$^{a}$\footnote{m.panchenko@campus.lmu.de} 

\vspace{.6truecm}


{\em $^a$Arnold Sommerfeld Center for Theoretical Physics\\
Department f\"ur Physik, Ludwig-Maximilians-Universit\"at M\"unchen\\
Theresienstr.~37, 80333 M\"unchen, Germany}


{\em $^b$Max-Planck-Institut f\"ur Physik\\
F\"ohringer Ring 6, 80805 M\"unchen, Germany}

{\em $^c$Center for Cosmology and Particle Physics\\
Department of Physics, New York University\\
4 Washington Place, New York, NY 10003, USA}

\end{center}

\vspace{0.5cm}

\begin{abstract}
\noindent  
 
{\small 
 Analyzing some well established facts, we give a model-independent parameterization of black hole 
 quantum computing in terms of a set of macro and micro quantities and their relations. These include the relations 
 between the extraordinarily-small energy gap of black hole qubits and important time-scales of information-processing, such as, scrambling time and Page's time.  
 We then show, confirming and extending previous results, that  other systems of nature with identical quantum informatics features are 
 attractive Bose-Einstein systems at the critical point of quantum phase transition. Here we establish a complete isomorphy 
 between the quantum computational properties of these two systems. In particular, we show that the quantum hair 
 of a critical condensate is strikingly similar to the quantum hair of a black hole.  Irrespectively whether one takes the
 similarity between the two systems  as a remarkable coincidence or as a sign of a deeper underlying connection, the following 
 is evident.  Black holes are not unique in their way of quantum information processing and we can manufacture 
 black hole based quantum computers in labs by taking advantage of quantum criticality. }

\end{abstract}

\thispagestyle{empty}
\clearpage

\section{Black Hole Qubits} 

 In general, for describing the processing of quantum information one uses the system of qubits. 
An elementary qubit represents a quantum system with two basis states that can be denoted by 
$|0\rangle$ and $|1\rangle$ respectively.   Here $0$ and $1$ refer to the eigenvalues of a particle number 
operator, $\hat{n}_b \equiv \hat{b}^{\dagger} \hat{b}$, of a quantum degree of freedom with corresponding creation 
and annihilation operators  $\hat{b}^{\dagger},  \hat{b}$.   
  
   The important characteristics of the qubit are:  The energy gap $\Delta E$ between its quantum states $|0\rangle$ and $|1\rangle$ and the interaction 
   strength, $\alpha$, of the qubit degree of freedom $\hat{b}$. 
     Obviously, the energy gap $\Delta E$ measures the energy cost 
   of information-storage in a given qubit $\hat{b}$.  This is the energy needed for bringing an isolated  qubit  from its ground-state  $|0\rangle$  into  the excited state  $|1\rangle$, 
  \begin{equation}
   \Delta E \, = \, E_1 - E_0 \, . 
  \label{delta10}
  \end{equation}
In the absence of interaction, $\alpha = 0$, the two states are energy eigenstates, and the information can be stored in the qubit permanently. 
 The quantity  $\alpha$ controls the time-scale of this information-storage. 
  Due to the interaction with other degrees of freedom, the information stored in the qubit sooner or later shall leak
  out.  
   The typical time-scale of this leakage ($\tau$) scales as,
   \begin{equation}
  \tau^{-1}   \sim {\Delta E \over \hbar}\alpha^{2} N_{qubit} \, , 
  \label{time-scale} 
  \end{equation}
  where $N_{qubit}$ is an effective number of degrees of freedom interacting with the qubit $\hat{b}$. 
   Their role can be played  by other qubits of the system and/or by the quanta of external radiation \footnote{In case when  $\hat{b}$ interacts with external degrees of freedom with different strengths, the above equation is generalized to, 
  $\tau^{-1}   \sim {\Delta E \over \hbar}\sum_j^{N_{qubit}} \alpha_j^{2}  \,$.}. 
   For example, after the time $\tau$ a qubit $\hat{b}$ can get de-excited due to a spontaneous emission of 
   a particle and the information stored in it leaks out in form of the emitted quantum.  Therefore, when talking about a well-defined qubit,  we shall implicitly assume  $\Delta E \gtrsim \hbar/\tau$.  
  
   Correspondingly, in a system of $N$ qubits, $\hat{b}_j, ~j=1,2,...N$,  the information can be stored in superpositions of 
  $2^N$ basic states, $|n_1,n_2, ... n_N\rangle$, where $n_j$ labels the eigenvalue of 
  the $j$-th qubit number operator, $\hat{n}_j$, and takes the two possible values, $n_j = 0$ or $1$.  \\
  
   Applying the above basic knowledge of quantum information to black holes, we discover that they exhibit extraordinary features. 
  It is well known \cite{Bent}  that black hole entropy scales as its area in Planck units,  $S = R^2/L_P^2$. Here 
  $R$ is the black hole radius and $L_P$ is the Planck length.  This fact implies that the number of qubits that store 
  black hole quantum information scales as $N \sim S$.   Let   $\hat{b}_j  ~(j=1,2,...N)$ be these qubits.  
  Then, each basic vector $|n_1, n_2, ... n_N\rangle$ (or their superposition) represents a black hole micro-state. They all 
  correspond to the same macro-state.  The total number of such basis states  is  $2^N$.  Of course, this self-consistently reproduces the correct scaling of entropy,  computed as log of number of states $S = ln (2^N) \sim N$.  \\ 
  
     The most remarkable thing about the black hole qubits  is their unusually-small energy-gap. Let us estimate it. In order to describe the same macro-state of a black hole, all the $2^N$ micro-states,  $|n_1,n_2, ... n_N\rangle$, must be crowded within the total energy gap $ \sim { \hbar \over R}$.  This means, that, at least for the majority of individual qubits, the energy gap must scale as 
\begin{equation}     
     \Delta E \, =\,  \epsilon {\hbar \over R}\,,~~ {\rm where}~~ \epsilon \sim {1 \over N} \, .
    \label{gap}
 \end{equation}   
   The importance of this result is that  it is not relying on any microscopic theory of a black hole.  Rather, it represents a parameterization that emerges from superimposing
 the expression of black hole entropy on the basics of quantum mechanics.  
 The parameter $\epsilon$, which we have introduced above, is a 
 very useful measure of the energy cost of information-storage in a given system
 and shall play an important role in what follows. \\

     Let us now try to appreciate the importance of this fact. For this, let us compare the energy gap of the black hole 
     qubit (\ref{gap}) with a typical gap in ordinary systems.   
  Consider a quantum system of size $R$ with weakly-interacting relativistic particles, e.g., 
  a massless quantum field in a box of size $R$.  The typical spacing between the energy levels in such a system 
  is set by ${\hbar  \over R}$.   Correspondingly,  the energy-cost  for the storage of a single information bit is $\Delta E \sim {\hbar  \over R}$.  This simple fact makes it clear how special the black hole qubits are. The qubit energy gap is 
   suppressed by an extra factor $\epsilon = {1 \over N}$, as compared to the expected $\sim {\hbar  \over R}$. 
 This is minuscule even for small  black holes. In order to 
   comprehend  the numbers, consider a black hole of earth's mass. Such a black hole has radius of order 
   centimeter, $R \sim $cm.  The corresponding number $N \sim 10^{66}$ and thus, $\epsilon \sim 10^{-66}$.  Now, for an ordinary quantum system 
  of  comparable size, the energy cost per single information bit would be $\Delta E \sim {\hbar \over R} \sim 10^{-5}$eV. 
   However, for a black hole qubit the analogous cost is only 
   $\Delta E_{BH}  \sim 10^{-71}$eV.  
       That is, the qubits of a cm-size black hole are  $66$ orders of magnitude cheaper than the energy gap expected from the minimal quantum uncertainty for  an ordinary system of similar size!   

 Another remarkable property of balck hole qubits is their extraordinarily weak coupling strength to other degrees
 of freedom. This is given by their gravitational coupling, $\alpha$,  which is equal to, $\alpha = {1 \over N}$.   \\  
 
   We would like to stress that none of the conclusions we have reached about black hole 
 qubits involved any assumptions or speculations about its microscopic structure. They simply followed from applying 
 the basic notions of quantum information to well-established  black hole properties, such as entropy. 
 
  The above raises the following legitimate question about the underlying microscopic physics:  \\
  
 {\it  How does a black hole manage to deliver qubits with enormously suppressed energy gap and weak coupling? }  \\  
     
      In this paper we shall capitalize on the answer provided by the theory \cite{cesargia, QC}.  According to this picture a
      black hole represents a multi-graviton state at the quantum critical point. The peculiar  features are explained by   
   the phenomenon that a gas of $N$ attractive bosons exhibits at the quantum critical point: The appearance of qubits with both the energy gap as well as the interaction strength suppressed  
   by powers of $N$, closely analogous to what we have observed for black hole qubits.  \\
   
    The present work will follow the road of trying to understand the 
 peculiarities of black hole quantum information in terms of  a much more general 
 natural phenomenon and using it for manufacturing black hole based computational sequence in labs.      
      
    The previous studies \cite{QC,nico,scrambling,gold,giamischa, mischa} have shown that, irrespectively of knowledge of microscopic black hole physics, there exist systems in nature that process information in a strikingly  similar way to black holes. These are attractive Bose-Einstein systems at the quantum critical point.  Virtually every known aspect of black hole information processing has a corresponding counterpart in a critical condensate.  
     
  In the previous work \cite{giamischa}  we have shown how, due to the universal property of appearance of energetically-cheap and weakly-interacting qubits,  the attractive Bose-Einstein systems at the quantum critical point can be used to realize black hole based quantum computing in labs.\\

 In the present paper, we shall further investigate the universal role of quantum criticality for  black hole type quantum 
 computing.  
 For making this role more transparent, we  give a model-independent parameterization of black hole 
 quantum information processing in terms of set of micro and macro parameters and relations among them.
 In this parameterization we shall not assume any particular microscopic theory and shall rely exclusively 
 on well-established black hole properties. 
  We shall then show that  this data is in one-to-one correspondence with analogous data obtained for 
  an attractive Bose-Einstein system at the quantum critical point, for which the microscopic theory is 
  very well defined and is fully operative.  \\ 
   
  As part of the quantum information analysis,  we shall study the coupling of the critical qubits to external fields, and  investigate mechanisms of information inscription  and retrieval.
  Remarkably, we discover that the behavior of the information-carrying hair of the critical condensate exhibits properties identical to black hole quantum hair. 
 In particular, the time-scale of quantum information retrieval is macroscopic 
 and is closely analogous to so-called Page's time\cite{page}.  This time-scale becomes infinite in the classical limit. Correspondingly, classically the hair of the critical condensate becomes infinite, but unresolvable within any finite time. \\ 
 
  Of course, in laboratory conditions one can manipulate the system externally. Correspondingly, information encoded 
  in a critical condensate can be decoded faster by taking the system away from the critical point.   We consider  examples of such computational sequences by  using external oscillators for dialing up the information in quantum state of critical qubits and then using these qubits as control logic gates.

\section{The role of Criticality for Quantum Information } 

 In order to understand the role of quantum criticality in appearance of energetically very cheap qubits, 
let us consider a gas of cold bosons in a $d$-dimensional torus of radius $R$  with a delta-function type attractive interaction.    
 This system  was used in \cite{QC} as the simplest prototype model for  imitating  
black hole information qubits at quantum criticality.
The Hamiltonian has the following form,     
   \begin{equation}
 {\mathcal H} \, = \, \int d^dx \, \psi^{+} \frac{- \hbar^2 \Delta}{ 2M} \psi \, - \, g \hbar  \ 
  \int d^dx \, \psi^{+}\psi^{+} \,\psi \psi \,, 
\label{Hnonderivative} 
\end{equation} 
where $\psi \, = \, \sum_{\vec{k}} \frac{1} {\sqrt{V}} {\rm e}^{i  {\vec{k} \over R} \vec{x}} \, a_k$, 
$V \, = \, (2\pi R)^d$ is the $d$-dimensional volume and $\vec{k}$ is the $d$-dimensional wave-number vector.  
$a_{\vec{k}}^\dagger, a_{\vec{k}}$ are creation and annihilation operators of bosons of momentum-number vector
${\vec{k}}$. These operators satisfy the usual algebra: $[a_{\vec{k}}, a_{\vec{k'}}^\dagger] = \delta_{\vec{k}\vec{k'}}$
and all other commutators zero. \footnote{For simplicity, below we shall not put hats on the operator symbols.}  
 The parameter $g$ controls the strength of the coupling. 
  The above system is known to exhibit a quantum critical behavior for certain values of the parameters. 
 For example,  in $d=1$, the quantum critical point marks the phase transition towards the bright soliton phase
(see, e.g., \cite{bright}).  
 
  We wish to trace how the black hole type qubits appear near the critical point. 
  For this, it is useful to represent Hamiltonian is  terms of momentum modes.  

  \begin{equation}
    {\mathcal H} \, = \, \Delta \left (  \sum_{k}  k^2 \, a_k^\dagger a_k  \, - \, {\alpha \over 4} \, \sum_{k_1+ k_2-k_3-k_4 = 0} 
   a_{k_1}^\dagger a^\dagger _{k_2}  a_{k_3}a_{k_4}\right) \,,    
     \label{Hamilton}
 \end{equation}
 where $\alpha \, \equiv \left ({g \over V R} \right) {2Rm \over \hbar} $  is a dimensionless coupling constant and 
 \begin{equation}
 \Delta \,  \equiv \, {\hbar^2 \over 2R^2 M}\, .
 \label{delta}
 \end{equation} 
   The latter quantity deserves special attention. In order to understand its physical meaning, let us 
  set $g=0$ while keeping all other parameters finite. Then, $\alpha=0$ and the interaction term vanishes. 
  The resulting Hamiltonian describes a gas 
  of free bosons in a box of size $R$.  In such a system the information can be stored in occupation numbers of momentum modes, $n_k$. 
  Since the particle number is conserved, the different choices of the total number $N$ represent super-selection 
 sectors.   Obviously, for a given $N$, the ground-state of the system corresponds to $n_0 = N$ and $n_k =0$ for all 
 $k\neq 0$.  
The lowest energy cost of storing a single bit of information  in such a system is given by the energy of $|k|=1$ one-particle state,  $n_1=1$. This energy cost is equal to $\Delta$.   
 
  We can now understand how bringing the system to a quantum critical point by attractive interaction 
  makes the energy gap for qubits cheaper.   For this let us switch on the attractive interaction by setting 
  $\alpha \neq 0$. The parameter that controls the proximity to criticality is $N\alpha$. 
   We shall start in the  regime $N\alpha \ll 1$. In this regime we can use Bogoliubov transformation for diagonalizing the Hamiltonian,

\begin{equation}
b_{\pm 1} = u a_{\pm 1} - v a_{\mp 1}^\dagger\,,
\end{equation}
where, 
\begin{equation}
u = \frac{1 + \epsilon}{2\sqrt{\epsilon}}\,,\hspace{2em}v = \frac{1 - \epsilon}{2\sqrt{\epsilon}}\,,  
\end{equation} 
with
 \begin{equation}
\epsilon \, = \, \sqrt{1-\alpha N} \,.
\label{eq:bogen}
\end{equation}

We thus arrive at the following effective low energy Hamiltonian for Bogoliubov modes:
 \begin{equation}
 {\mathcal H}_b  \,  = \,  \Delta \left ( \epsilon (b^\dagger_{-1} b_{-1} +  b^\dagger_{+1} b_{+1} )  \, + \, {1 \over N \epsilon^2} {\mathcal O} (b^4) \, \right ) \,, 
 \label{effective} 
 \end{equation}
 where ${\mathcal O} (b^4)$ stands for momentum-conserving quartic interactions among $b_{\pm1}$ and $b^\dagger_{\pm1}$ operators. 
 
  One thing that is immediately clear from the above expression is that by varying the parameter $\epsilon$ 
  we can make the energy gap  for  $b$-quibts {\it arbitrarily} small. Simultaneously, we can choose 
  $N$ sufficiently large, so that the effective Hamiltonian (\ref{effective}) is valid.  
  In turn, for any fixed values of $\epsilon$ and $N$, there is a maximal occupation number of $b$-modes,  above which the description in terms of the effective Hamiltonian breaks down.  This maximal occupation number is 
  $n_{max} \sim \epsilon^3N$.  Below we shall stay within this domain.   
   We shall assume that parameters of the system are chosen in such a way that 
  $n_{max}$ is sufficiently large.   For example, choosing  $\epsilon \sim N^{-1/4}$, we have 
  $n_{max}  \sim N^{1/4}$.    For $n_b \ll n_{max}$, the  interaction term can be ignored and 
  the system is well-described by the free $b$-modes over a sufficiently long time,  the duration of which is controlled by  $N$.  This was explicitly demonstrated 
by previous studies  (see \cite{giamischa} and references therein). 
  
    Notice that by taking variations of the model  (\ref{Hnonderivative}), we can easily change the number and the coupling 
   strength of the $b$-modes (see \cite{gold}) maintaining the key feature of appearance of the gapless weakly-interacting 
   $b$-qubits around the critical point.  Therefore, for further analysis we shall extract only this general property that both $\epsilon$ and $1/n_{max}$ are 
 small and scale as some inverse powers of $N$.    
   We shall specify the values for $\epsilon$ and $n_{max}$ only when applying our results to concrete systems, e.g., to black holes.    \\

    Thus, the important message, already much appreciated in the previous work \cite{QC, nico, scrambling, gold, giamischa}, is that the Bogoliubov qubits near the quantum critical point in the system of $N$ attractive bosons 
  carry features extremely similar to black hole quantum qubits, described in the introduction. 
  
  The very nature of this phase transition tells us that this remarkable coincidence is not an accident.  Indeed, the essence of the quantum phase transition leading to nearly-gapless quibits can be expressed as, 
    \begin{equation}
      (N~ {\rm delocalized~free~bosons})_{\alpha N < 1}  \rightarrow ({\rm localized~boundstate})_{\alpha N > 1}  \, .  
  \label{transition} 
   \end{equation} 
 It reflects the fact that the point $N = \alpha^{-1}$ marks the threshold beyond which the $N$ attractive bosons 
 prefer to bind together.  This immediately rigs a bell, because in gravity 
 $N = \alpha^{-1}$ is exactly the required number of gravitons, of any given wavelength $R$, that are able to form a black hole\cite{cesargia}.  Of course, $\alpha$ has to be understood as the gravitational coupling evaluated for the given  wavelength $R$, 
 i.e., $\alpha = {L_P^2 \over R^2}$. Again, this simple estimate is independent of a particular microscopic 
 model of a black hole and solely relies on the attractive nature and strength of gravitational interaction.  
  \\

   For a generic gas of attractive bosons, the precise nature of the  bound-state,
   formed for $\alpha N > 1$ is model-dependent. For example, in the model 
  (\ref{Hnonderivative})  for $d=1$,  in the regime  $\alpha N >1$, the bosons form a stable  bound-state, which represents a bright soliton.  For higher dimensional cases, the similar bound-state is unstable with respect to collapse.  These particularities, however, are not important for our present analysis, since in our discussion we shall stay in the vicinity of the critical point, around which the behavior exhibits some universal features.  This universality is the power of quantum criticality, which enables us  to reproduce the typical behavior of black hole qubits in other Bose-Einstein systems, in which the information can be stored cheaply and for a {\it macroscopically-long} time.  

\section{Quantum Informational Correspondence between  Critical Condensates and Black Holes}

 In this section we shall establish similarity, from a quantum information 
 point of view, between a critical Bose-Einstein condensate and 
 a black hole.  
 
   Both systems of interest are characterized by sets of macroscopic and microscopic parameters and the relations among these parameters. The term {\it similarity} means that there is a consistent mapping 
   between the parameters of the two systems.   Let us now review these parameters and the correspondence between them.\\

  For a critical Bose-Einstein system, such as given by (\ref{Hnonderivative}),  the macroscopic parameter is $N$, the occupation number of attractive bosons.  This parameter  simultaneously sets the  value of the total binding energy of the system at the quantum critical point: 
 \begin{equation} 
  E_{crit} = - N\Delta \, , 
 \label{bindingcond}  
 \end{equation} 
 where $\Delta$ is given  by (\ref{delta}).
   Because of quantum criticality, the same parameter $N$ also determines the values of the relevant microscopic parameters.  These are: The interaction strength among the bosons, $\alpha = {1\over N}$, and the suppression factor  for the energy gap of qubit, $\epsilon \sim {1 \over N^a}$.  Here $a$ is a model-dependent positive number (see the previous section). 
For the condensate described by (\ref{Hnonderivative}), 
within the validity of  Bogoliubov approximation, it takes values ${1\over 4} \leqslant a \leqslant {1 \over 3}$.       
    The  crucial property, shared both by black holes and critical condensates is that $\epsilon \rightarrow 0$ 
   for $N \rightarrow 0$. That is, the qubits become gapless in large-$N$ limit.

   An important fundamental parameter is the Compton wavelength  of the boson, $L_{Compt} \equiv {\hbar \over M}$, which sets the lower bound on $R$ for the validity of non-relativistic approximation.
 The convenient unit of energy is $\Delta$, given  by (\ref{delta}),  corresponding to the energy of the first non-zero momentum mode in the box of size $R$.\\
  
  Let us now review the analogous parameters for the black hole case. 
  The key macroscopic parameter of a black hole is $N \equiv {R^2 \over L_P^2}$. This parameter determines several things, in particular, the black hole entropy, mass and gravitational binding self-energy. 
   {\it A priory}, it is not related to any occupation number, but  it coincides with the 
   required number of gravitons of wavelength $R$ that can form a black hole.   
  This said, in our analysis we shall not make any assumption about the black hole's
 microscopic structure.  On the black hole side, we shall solely use the well established facts.  
   
  The reason why we are denoting the quantity $N$ by the same symbol as the number of bosons in the critical condensate is that this quantity  plays an identical role in defining 
 black hole's microscopic and macroscopic parameters. 

 Namely,  both the gravitational coupling, $\alpha = {1 \over N}$, 
 as well as the energy gap of black hole's qubits,  $\epsilon = {1 \over N}$, 
 are set by the inverse value of $N$, in a way that is very similar to the critical condensate. 
   The  black hole, just as the non-relativistic 
 condensate, also involves fundamental parameters.  
 This is the  Planck length, $L_P$, which 
sets the lower bound on $R$ for the validity of weak gravity description.  From this point of view, $L_P$ plays the role similar to 
the boson Compton wavelength $L_{Compt}$ in the non-relativistic condensate (\ref{Hnonderivative}). Both, 
 $L_P$  and $L_{Compt}$
 act as cutoffs 
of the respective effective descriptions.     
 Finally, the role of the energy unit $\Delta$ for the black hole case is given by the analogous quantity $\Delta_{BH} \equiv {\hbar \over R}$.  The difference between the two is the additional non-relativistic suppression factor, 
 ${L_{Compt} \over R}$, in case of the  condensate.   \\

 Quantity  $N$ also determines  the black hole's  macroscopic parameters. Notice, that 
 the black hole gravitational binding self-energy, is defined by $N$ and $\Delta_{BH}$ as, 
 \begin{equation} 
  E_{BH} = - N\Delta_{BH} \, , 
 \label{bindingcond}  
 \end{equation} 
which is identical to (\ref{bindingcond}).   
Another quantity determined by $N$ is the black hole half-life time, which 
in units of ${\hbar \over \Delta_{BH}}$, scales as  $t_{BH} \sim  N$. \\

 There are  two important time-scales in black hole information processing. First, according to Page\cite{page}, the time $t_{Page} \sim  N$ is required for the 
 start of efficient recovery of the black hole information. 
 The second scale is the information-scrambling time that was conjectured in 
 \cite{preskill} to scale as $t_{scramb} \sim  ln(N)$.  \\

  Interestingly, in the previous work \cite{scrambling} it was already 
 demonstrated that the critical Bose-Einstein system described by (\ref{Hnonderivative}) scrambles information in time $t_{scramb} \sim  ln(N)$. Further, in \cite{giamischa}, the time-scale for growth of entanglement was shown to scale as power-law in $N$.  
 
   Below, we shall establish yet another link by showing that the time-scale for resolving the quantum hair of the critical condensate scales as $t_{hair} \sim N$, in full correspondence with Page's time.  \\
 
  The correspondence between various characteristics of critical condensates 
  and a black hole is summarized in table \ref{BH}.  Viewed from this perspective, 
  the similarity between the two systems is truly remarkable. 
 \begin{table}
  \begin{center}
 \begin{tabular}{|l | l|}
  \hline
  \textbf{Critical Condensate} & \textbf{Black Hole}\\ 
  \hline
  Number of  Bosons $N$ & Number of Gravitons N \\
  \hline
  Coupling  $\alpha = {1 \over N}$ & Coupling  $\alpha = {1 \over N}$\\
  \hline
    Binding Energy $E_{crit} = - N\Delta$ & Binding Energy $E_{BH} = - N\Delta_{BH}$ \\
   \hline Energy unit $\Delta \equiv {\hbar L_{compt} \over R^2} $ &  Energy unit $\Delta_{BH} \equiv {\hbar \over R} $ \\
    \hline Cutoff  $L_{Compt}$ & Cutoff  $L_{P}$\\
     \hline  Scrambling time $t_{scramb} \sim ln(N)$ & 
     Scrambling time $t_{scramb} \sim ln(N)$\\
     \hline  Hair resolution time  $t_{hair} \sim N$ & 
   Hair resolution time  $t_{Page} \sim N$\\
       \hline  Qubit gap ${1 \over N^{a}} \Delta$ &  Qubit gap ${1 \over N} \Delta_{BH}$ \\
       \hline
\end{tabular}
  \end{center}
\caption{Correspondence between the micro and macro characteristics of 
a non-relativistic critical condensate and a black hole. 
The exact value of parameter $a$ is model-dependent, but it is universally positive, which 
ensures the black-hole-type suppression of the qubit energy gap for the critical condensate.  
 For the Hamiltonian given by (\ref{Hnonderivative}) 
within the validity of  Bogoliubov approximation it takes values ${1\over 4} \leqslant a \leqslant {1 \over 3}$. 
For Bosons with momentum-dependent coupling we  have $a=1$ \cite{gold}.   
}\label{BH}
\end{table}

\section{Readout of Information via Couplings to External Fields}
In order to gain further insight about the evolution of the
cheap qubits at quantum criticality both for real black holes, as well as for systems in laboratory, we need to understand 
the effect of their couplings to the external sources. 
Therefore, we shall investigate two possible simple couplings of the $b$-modes to an external single-frequency oscillator 
$c$.  This oscillator serves as a prototype for an electromagnetic mode in a laboratory cavity or a near-horizon electromagnetic field (or external  gravitational radiation) surrounding a real black hole. These couplings enable to encode information in the critical system and to read it out at some later time. We shall focus on this computational sequence in the next section.  
 We will first introduce the couplings and compute the time evolution of the coupled systems.
 For both couplings we study the consequences for the evolution of radiation modes. In this framework the properties of the external radiation depend on the internal state of the condensate.  
 Thus, by analyzing the former we can in principle gain knowledge about the latter.   
 
  Such analysis enables us to understand in simple terms the following: 

$~~~$
  
  {\it 1) How efficiently the external fields encode information into the qubits of the critical system; 
  
   and 
   
   2) The time-scale and the strength of the influence on external sources from   
   the internal quantum state of the critical qubits.  That is, the properties 
   of the quantum hair of the critical system.}
   
   $~~~$  
   
    It is surprising how far the analysis of the simple quantum system can take us.     
  We shall see that, as long as the coupling to external oscillators is not disturbing the quantum criticality, 
   the information-carrying hair of the critical condensate behaves strikingly similar to the black hole quantum hair.  \\

\subsection{The energetic coupling}
Let us first consider a Hamiltonian in which the external oscillator $c$ of frequency $\delta$ couples directly to the energy eigenstates of $b$, namely \cite{giamischa}
\begin{equation}
     H_{en} \, := \, \epsilon \,  b^\dagger b  \,  +  \nu\, b^\dagger b (c + c^\dagger) \, + \, \delta c^\dagger c \, ,
     \label{Hen}
 \end{equation} 
 where $\nu$ and $\delta$ are real parameters and the Hamiltonian is written in units of $\Delta=1$.
Let $\ket{\gamma}_c$ denote a coherent state of the radiation operator $c$, i.e. $c\ket{\gamma}_c=\gamma \ket{\gamma}_c$ and let $\ket{m}_b$ be the m-th eigenstate of $b^\dagger b$.  We can compute the time evolution of the states $\ket{m}_b\otimes \ket{\gamma}_c$ exactly, see \cite{giamischa}. The result is:
\begin{equation}\label{evHen}
 \ee^{-\iu\, H\, t} \ket{m}_b\otimes \ket{\gamma}_c= \ee^{-\iu t(\epsilon m -\frac{(m\nu)^2}{\delta})}\ket{m}_b \otimes \ket{\ee^{-\iu t \delta}\left(\gamma+\frac{m\nu}{\delta}\right)-\frac{m\nu}{\delta}}_c.
\end{equation}
The time evolution preserves coherence. Moreover,  the speed of the evolution depends on $\abs{\gamma+\frac{m\nu}{\delta}}$.  
Thus, the states with  
\begin{equation}                    
\gamma=-\frac{m\nu}{\delta}
\end{equation}
get ``frozen" and do not evolve, i.e., are energy eigenstates with energy $(\epsilon m -\frac{(m\nu)^2}{\delta})$. If $\gamma+\frac{m\nu}{\delta}\neq 0$, there will be fluctuations in the luminosity of $c$-radiation. From the above formulae follows:
\begin{equation}
\bra{m}\bra{\gamma} c^\dagger c\ket{m}\ket{\gamma} (t)=\text{const.}-\left(\gamma+\frac{m\nu}{\delta}\right)\frac{m\nu}{\delta}\cos(\delta t)
\label{energyfluctuation}
\end{equation}

 This  simple physical result has implications both for black-hole-based  quantum computing in the laboratory, as well as 
for understanding of the bare essentials of possible macroscopic astrophysical effects of the black hole quantum hair.

\subsection{The optical coupling}
 We will now consider a coupling analogous to the atom-field interaction in quantum optics, the system is a straightforward analogy of the Jaynes Cummings Hamiltonian. The total number of excited $b$ modes and photons is conserved so that the condensate can only emit photons and get de-excited or absorb photons and excite $b$ modes. The Hamiltonian is:
\begin{equation}\label{Hopt}
H_{opt}:=\epsilon \, b^{\dagger}b+(\delta+\epsilon) c^{\dagger}c+\frac{g}{2}(c \,b^{\dagger}+c^{\dagger}b).
\end{equation}
$H_{opt}$ is quadratic and can be diagonalized exactly so that time evolution is easy to compute. We are interested in the time evolution of coherent states. For that let us denote coherent states in $b$ and $c$ as $\ket{\vec{\gamma}}$ with $\vec{\gamma}=(\gamma_b, \gamma_c)$ and 
$$
b\ket{\vec{\gamma}}=\gamma_b\ket{\vec{\gamma}} \quad,\quad c\ket{\vec{\gamma}}=\gamma_c\ket{\vec{\gamma}}.
$$
Then we rewrite the Hamiltonian as
\begin{equation}
 H_{opt}= \begin{pmatrix}
  b^{\dagger}  && c^{\dagger}
\end{pmatrix}
M
\begin{pmatrix}
b\\
c
\end{pmatrix}
\end{equation}
with 
\begin{equation}
 M=
\begin{pmatrix}
\epsilon && \frac{g}{2} \\
\frac{g}{2} && \delta+\epsilon
\end{pmatrix}.
\end{equation}
It is not difficult to see that coherent states evolve as
\begin{equation}\label{evHopt}
\ee^{\,-\iu H_{opt} t}\ket{\vec{\gamma}}=\ket{\ee^{\,-\iu M\, t}\, \vec{\gamma}}.
\end{equation}
With this coupling one generally has fluctuations in the luminosity of external radiation $\bra{\vec{\gamma}}c^{\dagger} c\ket{\vec{\gamma}}$ depending on the value of the state $\gamma_b$. Namely:
\begin{equation}
\bra{\vec{\gamma}}c^{\dagger} c\ket{\vec{\gamma}}(t)=\abs{\gamma_c}^2\cos^2\left(\frac{\delta_g t}{2}\right)+\abs{\frac{g\,\gamma_b+\delta \,\gamma_c}{\delta_g}}^2\sin^2\left(\frac{\delta_g t}{2}\right),
\label{opticalfluctuation}
\end{equation}
where we have defined $\delta_g:=\sqrt{\delta^2+g^2}$.  These fluctuations of an external radiation mode reveal information about the
internal quantum state of  $b$-modes and thus
represent a {\it quantum hair}.  Below we shall see that when $b$-modes  are identified with Bogoliubov    
modes of a critical condensate, the quantum hair exhibits properties very similar to what one would expect for a black hole.

\subsection{Including more modes}
 Both above-considered cases can be straightforwardly generalized to 
 the situation when the external mode $c$ couples to several different species 
 of $b$-qubits.  
To cover the latter case we include $K$-number of species $b_j, ~ (j=1,2,...K)$ in  the Hamiltonians studied above and observe the change in the time evolution. The new Hamiltonians are
\begin{align}
H_{en,K}:&= \, \epsilon \,  \sum_{j=1}^K b^\dagger_j b_j  \,  +  \nu\,\sum_{j=1}^K b^\dagger_j b_j (c + c^\dagger) \, + \, \delta c^\dagger c \, \\
H_{opt,K}:&=\epsilon \,\sum_{j=1}^K b_j^{\dagger}b_j+(\delta+\epsilon) c^{\dagger}c+\frac{g}{2}\sum_{j=1}^K(c \,b_j^{\dagger}+c^{\dagger}b_j) 
\end{align}
The time evolutions are modified in the following way: For the energetic coupling states $\ket{m_1}_{b_1}\otimes \cdots \ket{m_K}_{b_K}\otimes\ket{\gamma}_c$ still evolve according to \ref{evHen} with $m$ replaced by $\sum m_j$, so there is no effective change. For the optical coupling there is no qualitative change either - concerning the time evolution of coherent states $H_{opt,K}$ is equivalent to coupling $c$ to a single mode  $b' \equiv  \frac{1}{\sqrt{K}} \sum_{j=1}^K \,  b_j$,
with the new coupling $g' \equiv  \sqrt{ K}g$ and all the other $K-1$-modes decoupled. 
Then all the previous results are the same.
We conclude that concerning the time evolution of condensate-hair in the couplings studied above there is no difference between the information
storage within low occupation numbers of many modes (like in a qubit system) or high occupation
numbers of few modes (like Bogoliubov modes). \\

\subsection{Similarity with Black Hole Quantum Hair}\label{hair}

 We  now wish to show that the above-derived evolution of the system of critical qubits coupled to an external radiation mode correctly reproduces a black hole type behavior.   
Namely, the internal 
quantum structure of the qubits of the critical condensate exhibits  {\it quantum hair}, which from the information point of view is very similar to the one of a black hole.

\subsubsection{Quantum Hair of Critical Condensate (Energy Coupling)} 

 The Hamiltonian (\ref{Hen}) is written for the generic values of 
 $\epsilon, \nu$ and $\delta$.  We shall now show that when the parameters are 
 such that the system of $b$-qubits maintains quantum criticality, 
 properties very similar to black hole quantum hair are reproduced. 
  In order not to affect criticality, the coupling between $b$ and $c$ modes 
  should not affect the gap of $b$-modes.  This is the case, if 
  $\nu \gamma \lesssim \epsilon$. 
  Then, according to (\ref{energyfluctuation}) the fluctuation in intensity and energy of external radiation is,  
 \begin{equation}
 {\Delta E \over E} \sim {m \nu \over \delta \gamma} \, \lesssim  \, {m \epsilon \over \delta \gamma^2}
 \label{fluctE1}
 \end{equation}
 where  (in our units)  $E \equiv  \delta \gamma^2$ is the radiation energy.   The time-scale on which this fluctuation 
 takes place is $t_{hair} = \delta^{-1}$.   This sets the minimal  time for resolving the internal state of 
 the condensate by measuring its influence on external radiation.    
 
 Let us first estimate $t_{hair}$ for the states corresponding to lowest energy levels of $b$-mode, 
 i.e., states with $m \sim 1$.   From (\ref{fluctE1}) it is then clear, that for triggering an order-one fluctuation even for the radiation of minimal intensity, i.e., 
 $\gamma \sim 1$, we need to take $\delta \sim \epsilon$. Thus, for an external observer the required time-scale for resolving the state of $b$-qubit, is $t_{hair} \sim \epsilon^{-1}$.   This relation is very similar to the one 
 between the black hole qubit gap and the information retrieval time.  
 For $\epsilon \sim {1 \over N}$, this gives Page's time
$t_{hair} \sim t_{Page} \sim N$.  \\  
 
  The key point here is that in (\ref{fluctE1})  $\gamma$ appears in the denominator.
 Consequently, in the classical limit, $\gamma \rightarrow  \infty$, 
 the fluctuation vanishes and the hair becomes {\it unresolvable}. 
  What we are observing is a manifestation of the {\it quantum hair } of the critical condensate strikingly 
  similar to the black hole quantum hair. \\

  Taking a larger occupation number $m$ is not  changing the above behavior,  since the relevant quantum states universally satisfy $m\epsilon  <  1$.  Then, it is clear that the time-scale $t_{hair} = \delta^{-1}$  on which the external  radiation is affected scales as the occupation number of quanta in this radiation, 
 $t_{hair} \sim \gamma^{2}$.  
Correspondingly, in the classical limit, when the external occupation number $\gamma^2$ diverges, the time-scale $t_{hair}$ also becomes infinite.  
This is exactly the essence of the quantum hair. \\  
 
  The above result nicely fits the outcome of the quantum 
$N$-portrait \cite{cesargia}  stating that black holes have quantum hair, which becomes infinite but unresolvable in the classical limit. \\

    If we modify parameters in such a way that the criticality, i.e., the gap of 
  $b$-qubits,  is compromised due to coupling to external radiation, the information retrieval   
 time can be shortened.  This situation can be achieved in laboratory condensates, but not in black holes.

\subsubsection{Quantum Hair of Critical Condensate for Optical Coupling} 
 
  The same result is obtained for the optical coupling (\ref{Hopt}).   Applying this case for modeling the coupling of a black hole qubit $b$ with external radiation
  $c$, we must take, 
 \begin{equation}
  \epsilon \sim {1 \over N} , ~~~g\sim \sqrt{\epsilon\delta} \sim  \sqrt{{\delta \over N}} \, .
  \label{choiseBHopt}   
\end{equation} 
The second choice makes sure that the gravitational nature of the mixing between the modes is taken into the account.  
 Studying the case  $\delta \ll {1 \over N}$ is meaningless, because 
 in this case the period of the fluctuation exceeds the life-time of a black hole we would like to model.  
 We can thus safely restrict ourselves to the case $g \lesssim \delta$.  
     
 For  $g \ll \delta$ the fluctuation of the intensity of the external radiation, described by equation (\ref{opticalfluctuation}), suggests that the intensity and energy fluctuate as 
  \begin{equation}
   {\Delta E \over E} \sim  {g\over \delta} 
 \left( {\gamma_b\over \gamma_c}  - {1 \over 2} {g \over \delta} \right )    
   \,.     
 \label{edeltaoptical1}
 \end{equation}
 Taking into account (\ref{choiseBHopt}) this can be written as 
  \begin{equation}
   {\Delta E \over E} \sim  {1\over \sqrt{N\delta}} 
 \left({\gamma_b \over \gamma_c}  - {\mathcal O}\left ({1\over \sqrt{N\delta}} \right) \right)    
   \,.     
 \label{edeltaoptical2}
 \end{equation}

  Let us consider some possible choices of $\delta$. For  $\delta \sim 1$, 
  the energy of the external radiation quantum is comparable to 
  a Hawking quantum of a prototype black hole, i.e., $\sim {\hbar \over R}$.   
  The fluctuation can be maximized by maximizing $\gamma_b$, which amounts to taking $\gamma_b \sim 
 \sqrt{N}$.  We then get  $ {\Delta E \over E} \sim  {1\over \gamma_c}$.  
  Thus, only a quantum external radiation, i.e., $\gamma_c \sim 1$, can fluctuate significantly.  Once again, we observe that the hair of the critical condensate, just like a black hole hair, is {\it quantum}.   
   
   Let us see what happens in the classical limit. Taking an extreme case $\delta \sim 1/N$,  the fluctuation can be made order one for arbitrarily large choice of $\gamma_c$.  E.g., we can take $\gamma_c \sim N$, in which case the energy of the radiation becomes comparable to black hole mass, but of course, the time-scale of fluctuation is  comparable to the black hole life time.    
   
It follows that such a significant 
fluctuation can only be observed over a time scale $\sim N$, which is analog of the black hole half-life time, 
or equivalently, of the order of Page's time $\sim N$.   Since, the classical limit corresponds to 
$N \rightarrow \infty$,  in this limit no effect can be detected on any finite time-scale. \\ 

  Thus,  the property that information encoded 
 in black hole micro-states cannot be retrieved on time-scale shorter than $\sim N$ is fully shared by other critical 
 systems.  
 
 Of course, in laboratory systems one has much more flexibility because the couplings can be manipulated
 externally and switched on and off according to the needs.  For example, there is no {\it a priory} restriction 
on the coupling $\nu$ which can be chosen suitably to the experiment. By tuning the coupling $\nu$ an experimentalist can move between the classical and quantum hair regimes.

 \section{Quantum Computational Sequence}
 
  As we have pointed out in the previous work \cite{giamischa},  the cheap Bogoliubov qubits appearing near quantum criticality can be used for performing a computational sequence.  By coupling these qubits to external degrees of freedom,  we can first encode information in them and later decode this information by using  the Bogoliubov mode as a control qubit.   We shall discuss these stages separately 
   
  \subsection{Dialing up  Mechanism: Evolution with coherent states} 
  
  In order to encode information in $b$-modes, we have to bring them to a particular state. 
  We can achieve this by coupling them to external degrees of freedom. Let us again denote this degree of freedom by 
  $c$.   The most convenient dialing mechanism depends on the particular form of the coupling between 
  $b$ and $c$ modes.  For simple forms of the couplings it is possible to bring the $b$-mode to a particular state 
  by time-evolving it from an initial state in which the state of the $c$-mode is close to being classical, i.e., is described by a coherent state. 
    We have shown \cite{giamischa} that such a situation allows in an experimental setup to dial-up the qubits of a quantum critical substance
 by using a certain resonance effect with an external classical field.  
 This process imitates the encoding of information in a black hole state through its interaction with an external classical radiation. We will now show how to dial particular $b$ states using the two couplings described above.
 
 \subsubsection{Dialing in the energetical coupling}
 
  This way of dialing was introduced in\cite{giamischa}. 
 From the time evolution as governed by $H_{en}$ follows, that a generic initial state  $\ket{in}_b \equiv \sum_m \beta_m\ket{m}_b$ (where $\beta_m$ are parameters) tensored with a coherent state $\ket{\gamma}_c$  will time evolve to a state dominated by a number eigenstate $\ket{m_0}_b$
 with   $m_0 = - \frac{2\delta \gamma}{\nu}$, provided that $\beta_{m_0}\neq 0$.
 
 Correspondingly we can dial-up the $b$-mode to a state dominated by $\ket{m_0}_b$ by choosing the coherent state  
state parameter of $c$ mode appropriately, i.e.  $\gamma=-\frac{m_0\nu}{2\delta}$. 

For  $\nu\gg\delta$, after evolving an initial state, 
 $\ket{in}_{bc} \equiv \ket{in}_b \otimes \ket{-\frac{m_0\nu}{2\delta}}_c$ for a sufficiently long time, 
only the $b$-mode state $m=m_0$ survives.  The contributions of $m \neq m_0$ time-averages to exponentially small values.  This is clear by probing the time-evolved initial state by a state  
$\ket{X}_b \otimes \ket{-\frac{m_0\nu}{2\delta}}_c$, where $\ket{X}_b$ is an arbitrary state ob $b$ having a non-zero overlap with  $\ket{m_0}_b$.  
  Then the matrix element
\begin{equation}
 \left(_c\bra{-\frac{m_0\nu}{2\delta}}\otimes\, _b\bra{X}\right) e^{-iHt}  \left(\ket{in}_b\otimes  \ket{-\frac{m_0\nu}{2\delta}}_c\right)
\end{equation}
is dominated by the contribution from $\ket{m_0}_b$, since 
\begin{equation}
 \left | _c\bra{-\frac{m_0\nu}{2\delta}} \left. \ee^{-\iu t 2\delta}\frac{(m_j-m_0)\nu} {2\delta}-\frac{m_j\nu}{2\delta} \right>_c \right |^2 =\ee^{-\frac{\nu^2}{\delta^2}(m_j-m_0)^2\sin^2(\delta t)}
\end{equation}
is exponentially suppressed for $\nu\gg\delta$ for any $m\neq m_0$, when averaged over time.
Thus, using coherent states of $c$ we can dial-up $b$-modes by collapsing randomly generated states 
to some fixed eigenstates of $b$.

\subsubsection{Dialing in the optical coupling}

With the optical coupling we can dial coherent states of $b$ (which will then remain coherent after the coupling is turned off) directly from the ground state. This follows from the time evolution formula (\ref{evHopt}). One simply couples the ground state of the condensate $\ket{\gamma_b=0}$ to a coherent state $\ket{\gamma_c}$ and obtains:

\begin{equation}
\ee^{\,-\iu H_{opt}\, t}\ket{ \begin{pmatrix}
0\\
\gamma_c
\end{pmatrix}}=:\ket{\begin{pmatrix}
\gamma_b(t)\\
\gamma_c(t)
\end{pmatrix}}
\end{equation}
with 
\begin{equation}
\gamma_b(t)= -\iu\,\ee^{\,-\iu \frac{t}{2}(2\epsilon+\delta)}\,\gamma_c \,\frac{g}{\delta_g}\sin \left(\frac{t\delta_g}{2}\right).
\end{equation}
Thus, essentially every coherent state of $b$ can be dialed from the vacuum by using the right external mode $\gamma_c$.

\subsection{$b$-Mode as Control Qubit} 
 
 In the previous work \cite{giamischa} we have shown that $b$-modes can act as control qubits.   We shall briefly review the example 
 considered there and then generalize it to different logic gates. 
 For this we again couple them to some external qubit $c$.  For this purpose it is enough to 
 take $c$ as a two-level system described by the states $|0\rangle_c$ and $|1\rangle_c$.

 Let us choose the form of the Hamiltonian (\ref{Hen}) (where we replace  $\delta$ by $2\delta$ for convenience) - analogous control gates can be designed using the coupling (\ref{Hopt}) as well.  The time evolution of states can be computed exactly.  
 We label the basic states as 
$$\left( \begin{matrix}
\alpha \\ 
\beta
\end{matrix} \right)_{m}
:= \ket{m}_b \left ( \alpha\ket{0}_c+\beta\ket{1}_c \right ) \, ,$$ 
and find
 \begin{equation}
 H \left(\begin{matrix}
 \alpha\\ \beta
 \end{matrix} \right)_m = A_m\cdot \left(\begin{matrix}
 \alpha\\ \beta
 \end{matrix} \right)_m,
 \end{equation}
    with $$A_m=\left(\begin{matrix}
    \varepsilon m && \nu m\\
    \nu m && \varepsilon m +2\delta
    \end{matrix}\right)\, , 
$$  
which can be easily exponentiated.  For  time-independent $\nu$, the time evolution of states is 
 \begin{align}\label{timevqub}
\left(\begin{matrix}
 \alpha\\ \beta
 \end{matrix} \right)_m(t) &= \ee^{-\iu t H}\left(\begin{matrix}
 \alpha\\ \beta
 \end{matrix} \right)_m = \nonumber \\
& \frac{\ee^{-\iu t (\delta +m\varepsilon )}}{\omega_m} 
 \left( \begin{matrix}
\alpha \, \omega_m  \cos (t \omega_m )+\iu \sin (t \omega_m ) (\alpha  \delta -\beta  m\nu) \\
 \beta \, \omega_m  \cos (t \omega_m )-\iu \sin (t \omega_m ) (\beta  \delta +\alpha m\nu)
\end{matrix}
\right)_m
\end{align}
where $\omega_m:=\sqrt{(m\,\nu)^2+\delta^2}$.

For time-dependent $\nu=\nu(t)$ one has to substitute $t\nu \rightarrow \int_0^t \nu(t') dt'$.

 We see that the Hamiltonian acts as a generalized controlled gate.

   For example, restricting to two levels $m=0,1$ in the limit $\delta=0$ and for $t\nu = \pi/2$ this system acts as a standard  CNOT gate. This gate is a controlled gate acting on two qubits (in our case $b$ and $c$).
   The first qubit, $b$, is playing the role of controller.  
 When this qubit is in the state $|0\rangle_b$, the CNOT gate  does nothing, whereas when the $b$-qubit  state is $|1\rangle_b$, gate performs a NOT operation on the second qubit: 
$\alpha\ket{0}_c+\beta\ket{1}_c \, \rightarrow \beta\ket{0}_c+\alpha\ket{1}_c$.

   It is clear from (\ref{timevqub})
   that Hamiltonian (\ref{Hen}) plays exactly this role for $\delta=0$ and for $t\nu = \pi/2$, because it acts trivially on the state with $m=0$, whereas for $m=1$ we have (up to a phase factor), 
   $$\left( \begin{matrix}
\alpha \\ 
\beta
\end{matrix} \right)_1
\rightarrow  \left( \begin{matrix}
\beta \\ 
\alpha
\end{matrix} \right)_1 \,.$$
   
   \subsubsection{ Example of three-qubit controlled gate: Toffoli gate}

      Higher order couplings of Bogoliubov qubits can easily  manufacture multiple qubit control gates. 
  As an example, we shall construct a Toffoli gate.   This gate is a three-qubit gate, with first two qubits acting 
  as control qubits for the third one.  The gate performs the NOT operation on the third qubit, only 
  when both of the first two qubits are in state $\ket{1}$.  Otherwise the gate does nothing.

   In our case the Bogoliubov Hamiltonian acts as Toffoli gate when the controlled qubit $c$ is coupled to the product of number operators 
  of two distinct Bogoliubov qubits,  say $b_{-1}$ and $b_{+1}$,  
  \begin{equation}
     H \, = \,  \epsilon (b^\dagger_{-1} b_{-1} +  b^\dagger_{+1} b_{+1} ) \,  +   
    \nu\, (b^\dagger_{-1} b_{-1}b^\dagger_{+1} b_{+1} ) (c + c^\dagger) \, + \, 2\delta c^\dagger c \, , 
     \label{Hamilton-b2}
 \end{equation} 
 where $\nu$ is again taken to be real. 
  The basic states now are labelled as, 
  $$\left( \begin{matrix}
\alpha \\ 
\beta
\end{matrix} \right)_{m_+m_-}
:=\alpha\ket{m_+}_{b_{+1}}\ket{m_-}_{b_{-1}}\ket{0}_c+\beta \ket{m_+}_{b_{+1}}\ket{m_-}_{b_{-1}}\ket{1}_c \, .$$ 
Correspondingly we find
 \begin{equation}
 H \left(\begin{matrix}
 \alpha\\ \beta
 \end{matrix} \right)_{m_+m_-} = A_{m} \cdot \left(\begin{matrix}
 \alpha\\ \beta
 \end{matrix} \right)_{m_+m_-},
 \end{equation}
    with $$A_m=\left(\begin{matrix}
    \varepsilon (m_+ + m_-)  && \nu m_+m_-\\
    \nu m_+m_- && \varepsilon (m_++m_-) +2\delta
    \end{matrix}\right)\, . 
$$  
Exponentiating it we get the following time evolution of states, 
 \begin{align}\label{timevqub}
\left(\begin{matrix}
 \alpha\\ \beta
 \end{matrix} \right)_{m_+m_-}(t) &= \ee^{-\iu t H}\left(\begin{matrix}
 \alpha\\ \beta
 \end{matrix} \right)_{m_+m_-} = \nonumber \\
& \frac{\ee^{-\iu t (\delta +(m_++m_-)\varepsilon )}}{\omega_m} 
 \left( \begin{matrix}
\alpha \, \omega_m  \cos (t \omega_m )+\iu \sin (t \omega_m ) (\alpha  \delta -\beta  m_+m_-\nu) \\
 \beta \, \omega_m  \cos (t \omega_m )-\iu \sin (t \omega_m ) (\beta  \delta +\alpha m_+m_-\nu)
\end{matrix}
\right)_{m_+m_-}
\end{align}
where $\omega_m:=\sqrt{(m_+m_-\,\nu)^2+\delta^2}$.

 It is obvious that for $\delta=0$ and $t\nu = {\pi\over 2}$ this evolution gives the Toffoli gate: 
 For  $(m_+,m_-)$ being $(0,0)$, $(1,0)$ and $(0,1)$ it acts on the state trivially, whereas 
 for $(1,1)$ we have, 
    $$\left( \begin{matrix}
\alpha \\ 
\beta
\end{matrix} \right)_{1,1}
\rightarrow  \left( \begin{matrix}
\beta \\ 
\alpha
\end{matrix} \right)_{1,1} \,.$$

\section{What Comes Next and What Can be Done?} 
There are two experimental directions in which one can go to test and extend the analogy between black holes and critical condensates. One is to construct a critical condensate in a lab,
couple it to external sources in some way, e.g., via the energetic or optical couplings as presented above, and study the system's information processing properties by encoding and decoding information \'a la sections 4 and 5. 
 The useful systems can be attractive cold atoms \cite{coldatoms} or magnons \cite{magnons}, which exhibit 
 properties discussed above.   I particular magnons can form droplets, which have been experimentally 
 studied \cite{magnet, magnet1}.

  When applied to astrophysical black holes,  our results suggest that it would be hopelessly hard to detect purely-gravitational quantum hair of a black hole by observing fluctuations in external radiation. However, this conclusion does not 
apply to the cases when the black hole hair is due to a large occupation number  $N_B$ of other particles, for example baryons. In such cases, the black hole 
hair can have a non-vanishing classical limit for $N \rightarrow \infty$ with $N_B/N =$ fixed \cite{baryonhair}. \\

Constructing an information theoretical analogue of a black hole in fact seems quite feasible, since, e.g., the Hamiltonian for magnons on a closed 1-dimensional wire with impurity coincides with the Hamiltonian of the attractive Lieb Liniger model
(\ref{Hnonderivative}), see e.g. \cite{magnons}, and therefore it could be a straightforward experimental setup on which the theory can be tested. Although damping makes the creation of magnetic droplets in realistic systems more complicated - in the experiments the damping has been compensated by spin-torque interactions \cite{magnet},\cite{magnet1} - the general features of a system that can undergo a quantum phase transition towards a solitonic phase should stay unchanged. In particular, right at the threshold current the appearance of nearly gapless states and the criticality of the ground state should manifest themselves in a qualitative way. There are several quantities that could be tested apart from the encoding/decoding information as presented above:

\begin{enumerate}
\item \textbf{The ground state correlation functions.} Before and after the phase transition, the ground state of the system can be regarded as classical to a good approximation, the lowest excitations can be described by Bogoliubov modes. At the critical coupling, however, the system is in the curious state on the edge of soliton formation. For the first time we now have access to an analytic expression for the critical ground state and for all correlation functions, see \cite{mischa}. By adiabatically bringing the magnon gas to the critical point and measuring the correlation functions, e.g. with the methods displayed in \cite{coupledgases}, one could test the predictions.

\item \textbf{Scrambling of Information in critical condensate.} 
By bringing the condensate into an overcritical regime and letting it evolve, one could measure the information scrambling time and test the prediction of \cite{scrambling}. There scrambling time was computed as a quantum break time and was shown to scale as $t_{scramb} \sim ln(N)$.

\item \textbf{Entanglement growth.} The critical point is characterized by high 
entanglement.  In \cite{nico} this was demonstrated numerically. 
 However, the time-scale for the growth is polynomial in $N$, which again points towards the existence of time-scales in critical condensates that are closely analogous to Page's time \cite{giamischa}.  The growth of the entanglement of a condensate at the critical dynamics has been calculated and a polynomial timescale was found analytically. By preparing the system to be a condensate and then bringing it to the critical coupling spontaneously (i.e., non-adiabatically), one could test the time evolution of correlation functions and measure the entropy growth.

\item \textbf{Classical and quantum hair.} As already mentioned in section \ref{hair}, by tuning the coupling one can go from a regime in which the condensate influences classical radiation to a regime in which only quantum radiation is significantly affected. Studying possible properties of the latter regime could give hints about observable consequences of quantum hair for real black holes.

\end{enumerate}

\section{Conclusions}

  The purpose of this work was to give further evidence for an information-theoretical
  correspondence between critical Bose-Einstein systems and black holes
   and to use this correspondence for manufacturing black hole based computational 
   sequences in laboratory conditions.    
  Quantum information connection between critical condensates and black holes  
  was originally suggested in\cite{QC} and investigated in previous papers \cite{QC,nico, scrambling, gold, giamischa, mischa} \footnote{Some new aspects of black hole quantum computing from different perspective will be considered in a complementary work \cite{benedikt}.}.  In \cite{giamischa} we have outlined a concrete black hole based quantum computational sequence in critical Bose-Einstein systems. \\
  
   In the present paper we have emphasized that the correspondence is a fact that is  independent of concrete assumptions about the microscopic physics of black holes. 
 We have shown that by superimposing the basic notions of quantum information with certain well-established properties of black holes allows us to give model-independent parameterization of information-theoretic properties of black holes in terms of a
 set of microscopic and macroscopic quantities and their relations.  We then showed 
 that this set of parameters and their relations are in one-to-one correspondence 
 with the analogous quantities of critical Bose-Einstein condensates.  
  On both sides these descriptions include:  The relations between the energy gap of qubits, their couplings and the important time-scales of the information-retrieval.
  For example, we have explicitly demonstrated that  quantum hair of a critical condensate has properties very similar to quantum hair of black holes. \\

 Irrespectively, whether one takes the above striking similarities as a simple coincidence
 or a sign of a fundamental universality of black holes and critical condensates, 
  it is clear that there exist systems of nature with black hole type properties of information-processing.  We can thus use the critical condensates for  black hole type quantum computations in laboratory. Extending the results of \cite{giamischa} we have designed examples of  computational sequences.        

\section*{Acknowledgements}

 We would like to thank Cesar Gomez for many valuable discussions.
The work of G.D. was supported by the Humboldt Foundation under Alexander von Humboldt Professorship, the ERC Advanced Grant ``UV-completion through Bose-Einstein Condensation (Grant No. 339169) and by the DFG cluster of excellence ``Origin and Structure of the Universe", FPA 2009-07908.

\end{document}